\documentclass[aps,prb,reprint,superscriptaddress,floatfix]{revtex4-2}

\usepackage{amsmath,amssymb,graphicx,bm}
\usepackage{hyperref}

\begin{document}

\title{Transconductance as a Probe of Valley Thermodynamics in Multilayer WSe$_2$} 

\author{Katsunori Wakabayashi}
\email{WAKABAYASHI.Katsunori@nims.go.jp}
\affiliation{Research Center for Materials Nanoarchitectonics (MANA),
National Institute for Materials Science (NIMS),
Tsukuba 305-0044, Japan}
\affiliation{Department of Nanotechnology for Sustainable Energy,
Kwansei Gakuin University, Sanda 669-1330, Japan}
\author{Souren Adhikary}
\affiliation{Department of Nanotechnology for Sustainable Energy,
Kwansei Gakuin University, Sanda 669-1330, Japan}
\author{Tomoaki Kameda}
\affiliation{Department of Nanotechnology for Sustainable Energy,
Kwansei Gakuin University, Sanda 669-1330, Japan}
\date{\today}

\begin{abstract}
Transconductance is a central figure of merit in field-effect transistors, typically governed by charge accumulation and carrier mobility.
In multilayer WSe$_2$ transistors, however, we show that the transconductance carries a nonlinear transport signature of inter-valley carrier redistribution between the $K$ and $\Gamma$ valleys.
This valley-crossover contribution suppresses transconductance in bilayer WSe$_2$ and reverses sign in trilayer, while remaining absent in single-valley systems.
Unlike extrinsic mechanisms such as trap-state filling or contact resistance, the anomaly does not affect the subthreshold swing within this mechanism and cannot be reproduced within conventional single-valley transport models.
Introducing the valley susceptibility $\chi_v \equiv \partial f_\Gamma/\partial V_{GS}$, bounded by an intrinsic thermodynamic limit $(4k_BT)^{-1}$, we quantify this response and show that it reaches ${\sim}0.20\,\mathrm{V}^{-1}$ in bilayer WSe$_2$ near threshold at room temperature.
The sign, magnitude, and temperature dependence of the anomaly provide directly measurable fingerprints of valley thermodynamics, establishing transconductance as an electrical probe of internal electronic degrees of freedom and revealing a previously unexplored nonlinear response in standard transistor measurements.
\end{abstract}
%
%

\maketitle

\section{Introduction}

Transconductance $g_m = dI_D/dV_{GS}$ is a central figure of merit in
field-effect transistors (FETs), quantifying the gate response of channel current
and governing amplifier gain and switching speed.
In conventional semiconductor devices, $g_m$ is determined by charge
accumulation and carrier mobility.
Its non-monotonic behavior is attributed
to extrinsic mechanisms such as trap-state
filling~\cite{Illarionov2020_Ditreview,Ghatak2015_trapping,Guo2015_trapping},
interface-induced mobility degradation~\cite{Cao2014_compactmodel}, or
contact resistance~\cite{Allain2015,Shen2021_contact}.

The transconductance of multilayer WSe$_2$ FETs,
however, carries a nonlinear transport signature of inter-valley carrier
redistribution.
When the $K$ and $\Gamma$ valence-band maxima are separated by an energy
comparable to $k_BT$, gate voltage not only accumulates charge but also
continuously redistributes holes between valleys with markedly different
effective masses~\cite{PaperI,Fallahazad2016,Movva2018}, generating a valley-crossover
contribution $g_{m,v}$ that suppresses $g_m$ in the bilayer (2L), reverses sign
in the trilayer (3L), and is absent in purely single-valley systems.

Non-monotonic behavior of $g_m$ is conventionally attributed to the
extrinsic mechanisms cited above, each of which simultaneously degrades the
subthreshold swing $SS \equiv \ln(10)\,dV_{GS}/d(\ln I_D)$
(the change in $V_{GS}$ required to change $I_D$ by one order of magnitude in the subthreshold regime).
In contrast, the valley-crossover contribution identified here is
qualitatively distinct.
It does not affect $SS$, reverses sign between bilayer and trilayer, and
\emph{cannot be reproduced within conventional single-valley transport models}.

The physical origin is the gate-induced competition between the light-hole
$K$ valley ($m_K^*\approx 0.40\,m_0$) and the heavy-hole $\Gamma$ valley
($m_\Gamma^*\approx 1.00\,m_0$)~\cite{Fallahazad2016,Movva2018,Xu2017_QHE} in multilayer WSe$_2$, whose relative
population is controlled electrically through gate-induced band
alignment~\cite{Zhu2011_SOC,Zhao2013_WSe2layers,Wickramaratne2014_indirect,Yeh2015_WSe2muARPES,Splendiani2010_MoS2PL}.
Shubnikov--de Haas spectroscopy on gated trilayer WSe$_2$ has
demonstrated this gate-controlled hole redistribution from $\Gamma$ to
$K$~\cite{Movva2018}.
However, its manifestation in the nonlinear transistor characteristics has
remained unexplored.

Previous valleytronics
studies~\cite{Schaibley2016_valleytronics,Mak2018_valleyreview,Xiao2007_valleyHall,Xiao2012_valley,Mak2010_MoS2direct,Mak2014_valleyHall}
focused on symmetry-related $K$--$K'$ valleys of
transition metal dichalcogenides (TMDCs)~\cite{Wang2012,Fiori2014_2Delectronics,Novoselov2016_2Dscience,Liu2021nature,Radisavljevic2011_MoS2,Desai2016},
exploiting spin-valley locking~\cite{Xu2014_spinpseudo,Riley2014_WSe2spinvalley}
and optical or magnetic control~\cite{Lai2023_valleytronic}.
Extending the equilibrium two-valley model of Ref.~\onlinecite{PaperI} to
the energetically nondegenerate $K$--$\Gamma$ regime, we introduce the
\emph{valley susceptibility} $\chi_v\equiv\partial f_\Gamma/\partial V_{GS}$
(where $f_\Gamma$ is the $\Gamma$-valley hole fraction)
and decompose the transconductance as $g_m = g_{m,0} + g_{m,v}$,
where $g_{m,0}$ describes the conventional gate response of charge accumulation
and $g_{m,v}$ describes the gate-driven inter-valley mobility shift.
We show below that $g_{m,v}\propto -\Delta\mu\,\chi_v$
($\Delta\mu \equiv \mu_K - \mu_\Gamma > 0$), where $\mu_K$ and $\mu_\Gamma$ are the mobilities of holes in the $K$ and $\Gamma$ valleys, respectively. The sign and magnitude of $g_{m,v}$ are governed by the thermodynamic
response of valley populations to gate voltage.

These results establish transconductance as a quantitative electrical probe of
valley thermodynamics in multi-valley semiconductors, providing
experimentally accessible signatures of internal electronic degrees of
freedom within standard transistor characterization.

\begin{figure*}[tb]
  \includegraphics[width=0.95\textwidth]{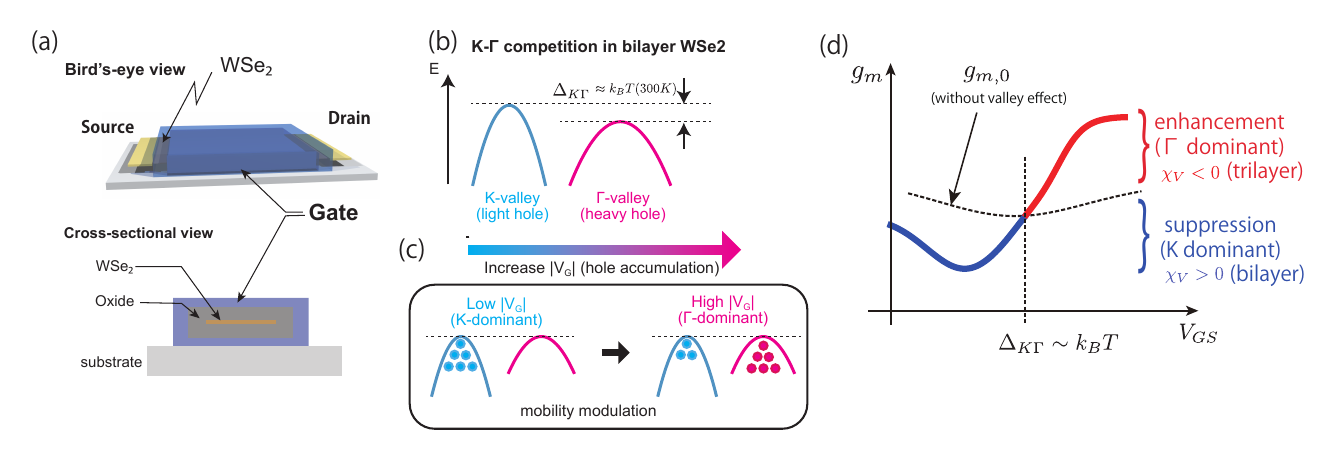}
  \caption{%
    (a) Schematic of the WSe$_2$ gate-all-around (GAA) field-effect
    transistor (bird's-eye and cross-sectional views).
    (b) Near-degenerate $K$ (light hole) and $\Gamma$ (heavy hole)
    valence-band maxima in bilayer WSe$_2$, separated by
    $\Delta_{K\Gamma}\approx k_BT$ at room temperature.
    (c) Increasing $V_{GS}$ accumulates holes and shifts the population
    from $K$-dominant to $\Gamma$-dominant transport, modulating the
    effective mobility.
    (d) Schematic $g_m$ vs.\ $V_{GS}$: the solid curve includes the
    valley-crossover contribution, while the dashed curve shows the
    reference without valley redistribution ($g_{m,0}$).
    The continuous crossover near $\Delta_{K\Gamma}\sim k_BT$ leads to
    suppression (bilayer) or enhancement (trilayer) of $g_m$.%
  }
  \label{fig:concept}
\end{figure*}

Figure~\ref{fig:concept} provides an overview of the physical picture.
The WSe$_2$ gate-all-around (GAA) FET geometry [panel~(a)] enables strong, symmetric electrostatic
control of the channel, making the $K$--$\Gamma$ valley occupation tunable by gate voltage.
In bilayer WSe$_2$ the light-hole $K$ and heavy-hole $\Gamma$ valence-band maxima are
separated by $\Delta_{K\Gamma}\approx k_BT$ at room temperature [panel~(b)], so both
valleys are thermally populated and their relative occupancy is gate-tunable.
As $V_{GS}$ increases, holes accumulate and progressively transfer from the high-mobility
$K$ valley to the low-mobility $\Gamma$ valley [panel~(c)], continuously reducing the
effective mobility.
Panel~(d) shows the resulting effect on $g_m$ schematically.
This valley-crossover contribution depresses $g_m$ below the charge-accumulation
baseline $g_{m,0}$ (dashed) in bilayer WSe$_2$, while the reversed sign of
$\Delta_{K\Gamma}$ in trilayer produces an enhancement.

\section{Two-Valley Model and Valley Susceptibility}

\subsection{Electrostatics and carrier statistics}
We model a WSe$_2$ GAA
FET~\cite{Mukesh2022_GAA_review,Tang2025_GAA2D,Ghosh2025_bilayerWSe2FET}
with gate oxide capacitance $C_{ox}$ per unit area and
interface-trap capacitance $C_{it}$.
Throughout this work, $V_{GS} \geq 0$ denotes the magnitude of the
applied gate voltage.
Increasing $V_{GS}$ corresponds to stronger hole accumulation in the
$p$-type channel.
The relation between gate voltage $V_{GS}$ and surface potential $\psi_s$ is
\begin{equation}
V_{GS} = \psi_s + \frac{Q(\psi_s)}{C_{ox}},
\qquad
Q = q\,p + C_{it}\psi_s,
\label{eq:VGS}
\end{equation}
where $q$ is the elementary charge and $p = p_K + p_\Gamma$ is the total
hole density.
Each valley contributes via the two-dimensional Fermi--Dirac
integral~\cite{Jimenez2012_driftdiff,Luryi1988_QC,Bennett2023_QC}
\begin{equation}
p_\nu = N_{2D,\nu}\ln\!\left(1 + e^{\eta_\nu}\right),
\quad
N_{2D,\nu} = \frac{m_\nu^* k_BT}{\pi\hbar^2}.
\label{eq:pnu}
\end{equation}
Here $\nu \in \{K, \Gamma\}$ labels the valley. The reduced chemical potentials
$\eta_\nu \equiv (E_v(\nu) - E_F)/k_BT$ ($E_F$: Fermi level) satisfy
$\eta_K = \eta_{K,0} + q\psi_s/k_BT$ and
$\eta_\Gamma = \eta_K - \Delta_{K\Gamma}/k_BT$,
where $\eta_{K,0}$ is the reduced chemical potential at the flat-band condition and
$\Delta_{K\Gamma} \equiv E_v(K)-E_v(\Gamma)$ is the $K$--$\Gamma$ valley
splitting (positive when $K$ is the valence-band maximum).

\subsection{Valley fraction and susceptibility}
We define the $\Gamma$-valley fraction
$f_\Gamma = p_\Gamma/p$ and the \emph{valley susceptibility}
\begin{equation}
\chi_v \equiv \frac{\partial f_\Gamma}{\partial V_{GS}}
= \frac{\partial f_\Gamma}{\partial\psi_s}\cdot\alpha_S,
\quad
\alpha_S = \frac{C_{ox}}{C_{ox}+C_Q+C_{it}},
\label{eq:chiv}
\end{equation}
where $C_Q = q\,\partial p/\partial\psi_s$ is the quantum capacitance and
$\alpha_S$ is the electrostatic screening factor.
In the nondegenerate (subthreshold) limit, where $\eta_K\ll 0$,
both valleys obey Boltzmann statistics
($p_\nu\propto e^{\eta_\nu}$) and $\partial f_\Gamma/\partial\psi_s = 0$
exactly.
The gate shifts $p_K$ and $p_\Gamma$ by the same multiplicative
factor $e^{q\delta\psi_s/k_BT}$, leaving their ratio---and hence $f_\Gamma$---unchanged.
The susceptibility $\chi_v$ is therefore a \emph{threshold phenomenon}.
It activates only in the degenerate regime ($\eta_K\gtrsim 0$), where the
Fermi--Dirac integral deviates from the Boltzmann approximation.
The threshold voltage $V_T$ is defined by the condition $\eta_K=0$
(onset of degeneracy).
The subthreshold regime corresponds to $V_{GS}<V_T$.

Three properties follow from Eq.~(\ref{eq:pnu}):
(i) in the Boltzmann limit, $f_\Gamma$ is governed by the dimensionless parameter
$x\equiv\Delta_{K\Gamma}/k_BT$ via
\begin{equation}
f_\Gamma = \frac{r_{\rm DOS}\,e^{-x}}{1+r_{\rm DOS}\,e^{-x}},
\quad
r_{\rm DOS} = m_\Gamma^*/m_K^*;
\label{eq:fGamma_x}
\end{equation}
(ii) in the full Fermi--Dirac regime near threshold, the activated
susceptibility obeys $\mathrm{sgn}(\chi_v) = \mathrm{sgn}(\Delta_{K\Gamma})$; and
(iii) the response is maximized at $x^* = \ln r_{\rm DOS}$, giving the
intrinsic bound
\begin{equation}
|\chi_v^{\rm max}| = \frac{1}{4k_BT}.
\label{eq:bound}
\end{equation}
This bound---not directly observable because $\alpha_S < 1$---serves as a
reference scale for device efficiency.
We distinguish $|\chi_v^{\rm max}|$ [Eq.~(\ref{eq:bound})], the unscreened
thermodynamic limit, from $|\chi_v^{\rm peak}|$, defined as the
gate-voltage maximum of the numerically computed $\chi_v(V_{GS})$ at
fixed $T$ and layer number with full electrostatic screening included.
For the WSe$_2$ GAA structure studied here
[equivalent oxide thickness $\mathrm{EOT}=0.7\,\mathrm{nm}$~\cite{Liu2012_ALD,Arutchelvan2023_process},
$C_{ox}\approx 49\,\mathrm{mF\,m}^{-2}$],
$|\chi_v^{\rm peak}|$ for 2L at $300\,\mathrm{K}$ is
$\approx 0.20\,\mathrm{V}^{-1}$, a factor of ${\approx}50$ below
the bound $|\chi_v^{\rm max}|=9.7\,\mathrm{V}^{-1}$, reflecting the suppression by
$\alpha_S\ll 1$ near threshold.

\begin{figure*}[tbp]
  \includegraphics[width=0.95\textwidth]{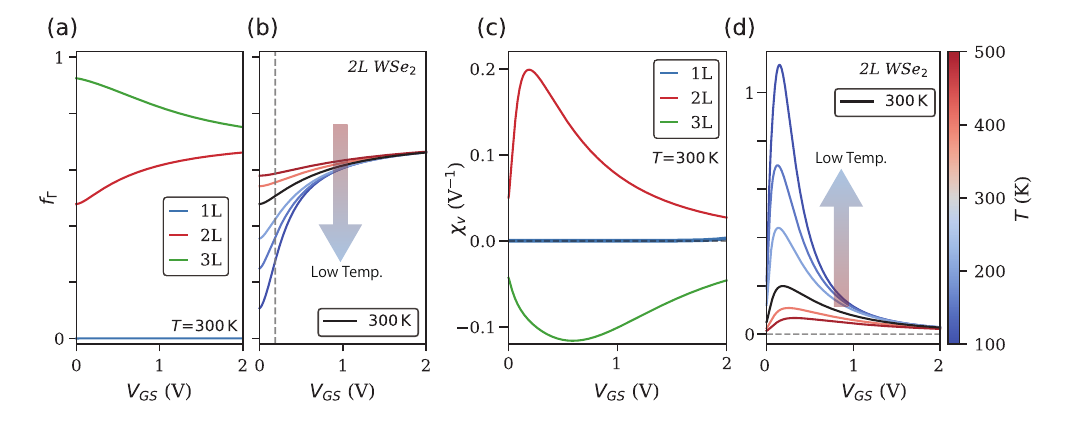}
  \caption{%
    $\Gamma$-valley fraction $f_\Gamma$ [panels~(a),(b)] and valley
    susceptibility $\chi_v = \partial f_\Gamma/\partial V_{GS}$
    [panels~(c),(d)] vs gate voltage for WSe$_2$ GAA FETs.
    (a),(c)~Layer-number dependence at $T=300\,\mathrm{K}$:
    bilayer (2L, $\Delta_{K\Gamma}\approx k_BT$) shows a positive peak
    $\chi_v^{\rm peak}\approx 0.20\,\mathrm{V}^{-1}$ near threshold;
    trilayer (3L) yields $\chi_v<0$, consistent with
    $\mathrm{sgn}(\chi_v)=\mathrm{sgn}(\Delta_{K\Gamma})$;
    monolayer (1L) is negligible.
    (b),(d)~Temperature dependence of 2L ($T=100$--$500\,\mathrm{K}$,
    blue to red; black: $300\,\mathrm{K}$):
    the peak $|\chi_v|$ grows with decreasing $T$, approaching the
    intrinsic $(4k_BT)^{-1}$ bound.%
  }
  \label{fig:fgamma_chiv}
\end{figure*}

Figure~\ref{fig:fgamma_chiv} shows $f_\Gamma(V_{GS})$ [panels~(a),(b)] and
$\chi_v(V_{GS})$ [panels~(c),(d)] computed for this device structure
(parameters in the Supplemental Material, Table~S6).
The layer-number dependence at $T=300\,\mathrm{K}$ [panels~(a),(c)] confirms the sign rule.
The bilayer ($\Delta_{K\Gamma}>0$) develops a positive
$\chi_v$ peak of ${\approx}0.20\,\mathrm{V}^{-1}$ near threshold as the $\Gamma$-valley
fraction grows with gate voltage, whereas the trilayer ($\Delta_{K\Gamma}<0$)
yields an opposite-sign peak of ${\approx}-0.10\,\mathrm{V}^{-1}$.
The monolayer (1L, $x\approx 19$) contributes negligible $\chi_v$ because the $\Gamma$ valley is
thermally inaccessible.
The temperature dependence of 2L [panels~(b),(d)] reveals the thermodynamic
origin of the effect.
As $T$ decreases, $|\chi_v^{\rm peak}|$ grows following
the $(4k_BT)^{-1}$ scaling of the intrinsic bound [Eq.~(\ref{eq:bound})],
in contrast to the thermally activated behavior expected for trap-state mechanisms.

\section{Transconductance Decomposition and Valley Anomaly}

\subsection{Effective mobility and drain current}
The $K$ and $\Gamma$ valleys carry hole mobilities $\mu_K$ and $\mu_\Gamma$,
with $\mu_K>\mu_\Gamma$ because $m_K^*<m_\Gamma^*$~\cite{Allain2014_WSe2mobility,Gunst2025_WSe2mobility,Movva2015_WSe2dualgate,Fivaz1967,Kormanyos2015_kp,Liang2023CPL}.
The valley-averaged effective mobility
\begin{equation}
\mu_{\rm eff} = \mu_K - \Delta\mu\,f_\Gamma,
\quad
\Delta\mu \equiv \mu_K - \mu_\Gamma > 0,
\label{eq:mueff}
\end{equation}
decreases as holes transfer into the heavier $\Gamma$ valley.
In the linear regime,
\begin{equation}
I_D = \frac{W}{L}\,\mu_{\rm eff}\,qp\,V_{DS},
\label{eq:ID}
\end{equation}
where $W$ and $L$ are the channel width and length, respectively,
and $V_{DS}$ is the drain-source voltage.

\subsection{Transconductance decomposition}
Differentiating Eq.~(\ref{eq:ID}) with respect to $V_{GS}$ and using
$\partial\mu_{\rm eff}/\partial V_{GS} = -\Delta\mu\,\chi_v$ yields
\begin{align}
g_m &= g_{m,0} + g_{m,v}, \label{eq:gm_decomp} \\
g_{m,0} &= \frac{W}{L}\,V_{DS}\,\mu_{\rm eff}\,C_Q\,\alpha_S,
\label{eq:gm0} \\
g_{m,v} &= -\frac{W}{L}\,V_{DS}\,\Delta\mu\,\chi_v\,qp.
\label{eq:gmv}
\end{align}
The charge-accumulation term $g_{m,0}$ is the standard transconductance.
The valley-crossover term $g_{m,v}$ is a contribution from the valley degree of freedom
whose sign is set by $\mathrm{sgn}(\chi_v)$:
\begin{equation}
\mathrm{sgn}(g_{m,v}) = -\mathrm{sgn}(\Delta_{K\Gamma}).
\label{eq:gmv_sign}
\end{equation}
For bilayer WSe$_2$ ($\Delta_{K\Gamma}>0$, $\chi_v>0$):
$g_{m,v}<0$ and gate-induced transfer of holes from high-mobility $K$ to
low-mobility $\Gamma$ \emph{suppresses} $g_m$.
For trilayer ($\Delta_{K\Gamma}<0$, $\chi_v<0$):
$g_{m,v}>0$ and the reverse transfer \emph{enhances} $g_m$
above the charge-accumulation baseline $g_{m,0}$.

\begin{figure*}[tbp]
  \includegraphics[width=0.95\textwidth]{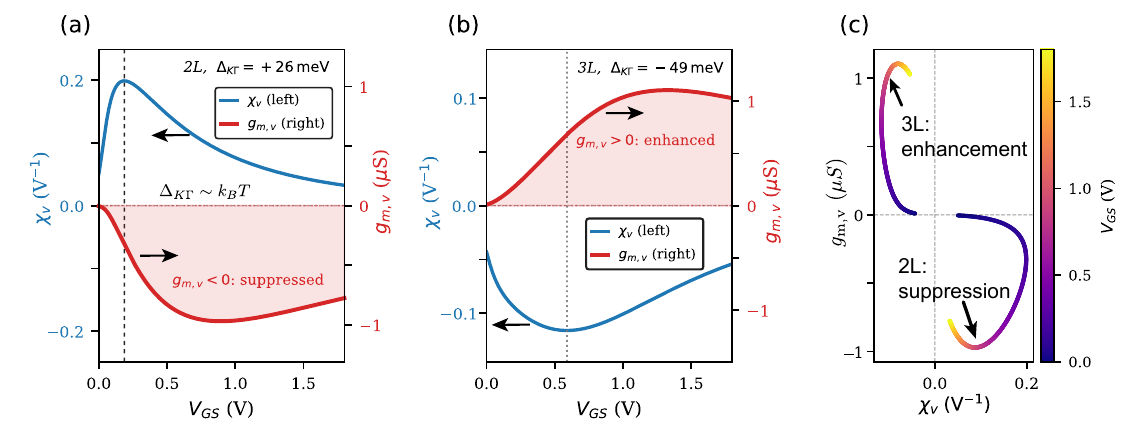}
  \caption{%
    Valley susceptibility $\chi_v$ and valley-crossover transconductance
    $g_{m,v}$ for WSe$_2$ GAA FETs at $T=300\,\mathrm{K}$,
    $V_{DS}=50\,\mathrm{mV}$.
    (a) Bilayer ($\Delta_{K\Gamma}=+26\,\mathrm{meV}$): $\chi_v>0$ (blue,
    left axis) drives $g_{m,v}<0$ (red, right axis), suppressing $g_m$.
    (b) Trilayer ($\Delta_{K\Gamma}=-49\,\mathrm{meV}$): the sign reversal
    $\chi_v<0$ yields $g_{m,v}>0$, anomalously enhancing $g_m$.
    Shaded regions indicate the sign of $g_{m,v}$; dotted vertical lines
    mark the crossover voltage $\Delta_{K\Gamma}\sim k_BT$.
    (c) Parametric correspondence ($\chi_v$,\,$g_{m,v}$) for both layers,
    with $V_{GS}$ encoded in color (plasma colormap, 0 to 1.8\,V).
    The two layers occupy opposite quadrants, demonstrating
    the sign rule $\mathrm{sgn}(g_{m,v})=-\mathrm{sgn}(\Delta_{K\Gamma})$.%
  }
  \label{fig:gm_combined}
\end{figure*}

A clear demonstration of the sign rule [Eq.~(\ref{eq:gmv_sign})]
is the parametric plot of $(\chi_v,\,g_{m,v})$ shown in
Fig.~\ref{fig:gm_combined}(c), where $V_{GS}$ is encoded in color.
The 2L trace occupies exclusively the fourth quadrant ($\chi_v>0$,
$g_{m,v}<0$) and the 3L trace the second quadrant ($\chi_v<0$,
$g_{m,v}>0$).
The two layers are confined to \emph{opposite} quadrants across the entire gate-voltage range.
This anti-correlation is a consequence of Fermi--Dirac valley thermodynamics
and cannot be reproduced within conventional single-valley transport models
or by interface traps and contact resistance.
The latter mechanisms degrade $SS$ simultaneously, whereas $g_{m,v}$
does not affect $SS$ within this model.

The individual gate sweeps are shown in Figs.~\ref{fig:gm_combined}(a)
and~\ref{fig:gm_combined}(b).
For bilayer [Fig.~\ref{fig:gm_combined}(a)], $\chi_v$ (blue, left axis)
peaks near $V_{GS}\approx 0.15\,\mathrm{V}$ at $\chi_v^{\rm peak}\approx
0.20\,\mathrm{V}^{-1}$, and $g_{m,v}$ (red, right axis) becomes increasingly negative
beyond threshold, reaching $g_{m,v}\approx -1.1\,\mu\mathrm{S}$ at its
minimum.
For trilayer [Fig.~\ref{fig:gm_combined}(b)], the sign is reversed
throughout: $\chi_v<0$ and $g_{m,v}>0$,
with a peak enhancement of $\approx +1.1\,\mu\mathrm{S}$.
These opposite-sign behaviors are consistent with the sign rule
[Eq.~(\ref{eq:gmv_sign})] and the freestanding-slab density functional theory (DFT) valley splittings
$\Delta_{K\Gamma}=+26\,\mathrm{meV}$ (bilayer) and
$-49\,\mathrm{meV}$ (trilayer) (Supplemental Material, Table~S1).

\subsection{Proximity of Bilayer WSe\texorpdfstring{$_2$}{2} to the Optimal Thermodynamic Condition}
The magnitude of $g_{m,v}$ is maximized when $|\chi_v^{\rm peak}|$ is
largest, which from Eq.~(\ref{eq:bound}) occurs at
$x^* = \ln r_{\rm DOS}$.
For 2L WSe$_2$ at room temperature and zero strain,
our first-principles DFT calculations [Perdew--Burke--Ernzerhof (PBE) functional with spin--orbit coupling (SOC) and van der Waals (vdW) correction, freestanding slab]
yield $\Delta_{K\Gamma} = +26\,\mathrm{meV}$ (Supplemental Material, Table~S1), giving
$\Delta_{K\Gamma}/k_BT \approx 1.0$ and
$\ln r_{\rm DOS}\approx 0.93$, placing the bilayer near the optimal
thermodynamic condition without applied strain or compositional modification.
The monolayer ($x\approx 19$) and trilayer ($x<0$) lie far from this optimal condition.
The 2L value $|\Delta_{K\Gamma}|=26\,\mathrm{meV}$ places the bilayer
close to the $K$--$\Gamma$ crossover, and the sign of $\Delta_{K\Gamma}$ is sensitive to the dielectric
environment.
Micro-angle-resolved photoemission spectroscopy ($\mu$-ARPES) measurements of bilayer WSe$_2$ transferred onto
a native-oxide-terminated Si substrate report
$\Delta_{K\Gamma}\approx -0.14\,\mathrm{eV}$~\cite{Yeh2015_WSe2muARPES},
reflecting substrate-induced dielectric screening and epitaxial strain, both of which are absent in the freestanding-slab DFT calculation and in the gate-dielectric-encapsulated device studied here.
More generally, the sign rule [Eq.~(\ref{eq:gmv_sign})] shows that
the sign of $g_{m,v}$ is determined solely by $\mathrm{sgn}(\Delta_{K\Gamma})$:
even if the K--$\Gamma$ crossover shifts to a different layer number in a
particular dielectric environment, the valley-crossover contributions
of opposite signs remain observable at the crossover layer,
and the mechanism is unchanged.
Biaxial strain shifts $\Delta_{K\Gamma}$
continuously~\cite{Johari2016_WSe2strain,Desai2014_WSe2strain,Yang2024_biaxialstrain,He2024_strainFET,Lau2024_straintronics,Ahn2017NatComm,Schmidt2024_straintransfer}
and provides experimental access to different regimes of $x$.
Compressive strain moves the 2L system slightly closer to $x^*$, while
tensile strain drives it toward the sign-reversal point
$\Delta_{K\Gamma}=0$~\cite{PaperI}.

\subsection{Distinguishing valley crossover from trap effects}
A key diagnostic criterion follows from the independence of $SS$ from
$g_{m,v}$.
The subthreshold swing
$SS = \ln(10)(k_BT/q)(1+(C_Q+C_{it})/C_{ox})$
is governed by electrostatics and is insensitive to the valley character
of the carriers.
In the subthreshold regime $C_Q\ll C_{ox}$ and $f_\Gamma$ varies
negligibly, so $\chi_v\approx 0$ and $g_{m,v}\approx 0$.
Interface traps degrade both $SS$ and the overall $g_m$ profile simultaneously.
By contrast, the valley-crossover anomaly $g_{m,v}$ remains observable even when $C_{it}$ substantially degrades $SS$.
Its sign, set by $\mathrm{sgn}(\Delta_{K\Gamma})$, is unaffected by trapping.
Its temperature dependence follows $(4k_BT)^{-1}$ rather than the
thermally activated behavior $\propto\exp(-E_a/k_BT)$ expected for
deep-level traps~\cite{Leonhardt2017_Dit,Ali2024_WSe2SS,Lee2022IEDM_MoS2SS}.
An anomaly in $g_m$ that
(i)~persists despite trap-induced degradation,
(ii)~reverses sign between 2L and 3L, and
(iii)~grows in amplitude with decreasing temperature
provides a robust experimental signature of valley-crossover origin.
The ${\sim}T^{-1}$ growth is opposite to the thermally activated
behavior expected for deep-level interface traps, making temperature
sweeps an experimental discriminant (Supplemental Material, Sec.~S9).

\section{Universal Scaling and Device Design}

\begin{figure}[tbp]
  \includegraphics[width=0.9\columnwidth]{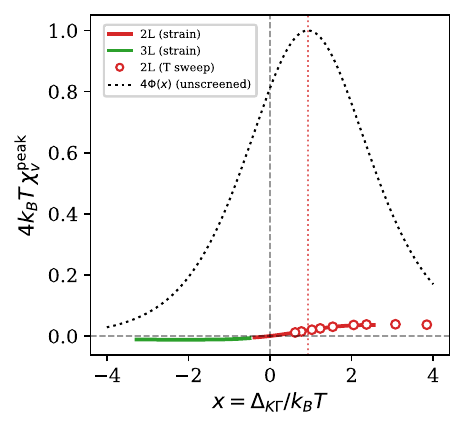}
  \caption{%
    Dimensionless valley susceptibility $4k_BT\chi_v^{\rm peak}$ as a
    function of $x=\Delta_{K\Gamma}/k_BT$.
    Solid lines: biaxial strain sweep at $T=300\,\mathrm{K}$ for 2L (red)
    and 3L (green).
    Open circles: temperature sweep at zero strain for 2L.
    Dotted curve: theoretical unscreened bound $4\Phi(x)$
    [Eq.~(\ref{eq:Phi})], which peaks at unity.
    All data lie ${\approx}50\times$ below the bound, reflecting
    the $\alpha_S\ll 1$ electrostatic screening penalty near threshold.
    The strain and temperature routes do not collapse onto a single curve
    because temperature simultaneously modifies $x$ and relaxes screening.
    Vertical markers: $x=0$ (sign reversal) and
    $x=\ln r_{\rm DOS}$ (optimal condition).%
  }
  \label{fig:universal}
\end{figure}

The intrinsic valley response is universal.
In the Boltzmann limit,
$\chi_v^{\rm peak}\propto\Phi(x)/(k_BT)$, where
\begin{equation}
\Phi(x) = \frac{r_{\rm DOS}\,e^{-x}}{(1+r_{\rm DOS}\,e^{-x})^2}
\label{eq:Phi}
\end{equation}
peaks at $\Phi_{\max}=1/4$ independent of $r_{\rm DOS}$.
Device-level screening breaks this universality.
Figure~\ref{fig:universal} shows $4k_BT\chi_v^{\rm peak}$ versus
$x = \Delta_{K\Gamma}/k_BT$, obtained by varying either biaxial strain
at $T=300\,\mathrm{K}$ or temperature at zero strain.
The two routes do not collapse onto a single curve---temperature
simultaneously tunes $x$ and reduces $C_Q\propto N_{\rm 2D}\propto T$,
relaxing the screening factor $\alpha_S$, whereas strain modulates only
$\Delta_{K\Gamma}$.
All data lie ${\approx}50\times$ below the theoretical bound $4\Phi(x)$ (dotted),
reflecting the electrostatic screening suppression by $\alpha_S\ll 1$ near threshold.
This non-universality is not a breakdown of the thermodynamic framework.
Rather, it quantifies the role of device geometry.
Increasing $C_{ox}$ (thinner or higher-$\kappa$ gate
dielectric~\cite{Tang2025_highk,Lin2025NatElectron,Radisavljevic2012_dualgate})
or operating at lower temperature is the most effective route to
approach the bound and amplify the observable valley anomaly.

\section{Discussion}

The transconductance decomposition [Eqs.~(\ref{eq:gm_decomp})--(\ref{eq:gmv})]
predicts three experimentally testable signatures:
(i) $g_{m,v}$ is absent in the monolayer but pronounced in the bilayer
and trilayer, with opposite signs;
(ii) $|g_{m,v}^{\rm peak}|$ grows as temperature decreases following
$(4k_BT)^{-1}$ scaling;
(iii) the anomaly persists under interface-trap degradation that suppresses
$SS$, providing an unambiguous valley-crossover criterion.
For a $W=L=1\,\mu\mathrm{m}$ device at 300\,K,
$|g_{m,v}^{\rm peak}|\approx 1\,\mu\mathrm{S}$, representing a 7\% anomaly
relative to $g_{m,0}$, growing to $\approx 17\%$ at 100\,K.
Numerical differentiation of a precisely measured $I_D$--$V_{GS}$ curve,
or lock-in detection of the ac transconductance,
provides sufficient sensitivity for detection.

The present analysis considers hole transport exclusively, which is
appropriate for the $p$-type operation regime studied here.
In the $p$-type accumulation regime ($V_{GS} > 0$), the Fermi level lies in the valence-band manifold and the
conduction-band electron population is exponentially suppressed.
Electron contributions become relevant in bipolar or ambipolar operation,
where an analogous $K$--$\Gamma$ valley-crossover term would appear in the
electron transconductance.
The analysis proceeds analogously with conduction-band effective masses.

The quasi-equilibrium treatment assumes that the intervalley scattering
time $\tau_{\rm iv}$ is much shorter than the carrier transit time
$\tau_{\rm tr}=L/v_d$~\cite{Sohier2018_mobility,Sohier2019_valleyEng,Afrid2026_intervalley,Bae2022_intervalley}.
When $\tau_{\rm iv}\gg\tau_{\rm tr}$, the valley population is frozen at
the source-contact value and $g_{m,v}$ is suppressed,
making the channel-length dependence of $g_{m,v}$ an additional
experimental observable.

The framework extends beyond WSe$_2$.
The essential requirement is that multiple valleys with distinct
effective masses are separated by $\Delta_{K\Gamma}\sim k_BT$.
Analogous physics is expected in other few-layer
TMDCs~\cite{Manzeli2017_TMDreview,Komsa2012_indirect,Ramasubramaniam2011,Godiksen2022_valley,Zhang2014_MoSe2,Geim2013_vdW},
including systems where electrical valley control has been demonstrated via
ferroelectric gating~\cite{Li2022_ferroelectric_valley},
and in conventional multi-valley semiconductors under uniaxial stress.
In each case, the transconductance serves as an electrical probe of
internal valley thermodynamics---one that requires no optical
or magnetic field, and is fully compatible with standard
transistor characterization~\cite{Das2021_2Dreview,Smets2023_WSe2improvements,Ghosh2025_bilayerWSe2FET,Pack2024_WSe2,Arutchelvan2023_process}.

\section{Conclusion}

We have shown that gate-induced redistribution of holes between the $K$
and $\Gamma$ valleys of WSe$_2$ produces a distinct contribution
$g_{m,v} = -\Delta\mu\,(W/L)V_{DS}\,qp\,\chi_v$ to the transconductance.
This valley-crossover term suppresses $g_m$ in bilayer and enhances it in
trilayer WSe$_2$, in both cases without affecting the subthreshold swing.
The valley susceptibility $\chi_v$ vanishes exactly in the subthreshold
regime, peaks near threshold with a magnitude bounded by $(4k_BT)^{-1}$,
and is suppressed by a factor of ${\sim}50$ due to electrostatic screening in
realistic GAA devices.
In our DFT calculations, freestanding bilayer WSe$_2$ at room
temperature satisfies the optimal thermodynamic condition
$\Delta_{K\Gamma}\approx k_BT\ln r_{\rm DOS}$.
These results establish the transconductance as a quantitative probe of valley
thermodynamics in multi-valley two-dimensional transistors---an
electrical diagnostic of internal electronic degrees of freedom
requiring no optical or magnetic field, accessible through standard electrical characterization.

\begin{acknowledgments}
This work was supported by JSPS KAKENHI (Grants No.~JP25K01609,
No.~JP22H05473, and No.~JP21H01019), JST CREST (Grant No.~JPMJCR19T1).
K.W. acknowledges financial support from the Sumitomo Foundation
(Grant No.~2401203).
\end{acknowledgments}

\section*{Data Availability}
The data generated and analyzed in this study are available from the
corresponding author on reasonable request.

\bibliography{references}

@article{Mukesh2022_GAA_review,
  author  = {Mukesh, S. and Zhang, J.},
  title   = {A Review of the Gate-All-Around Nanosheet {FET}
             Process Opportunities},
  journal = {Electronics},
  volume  = {11},
  pages   = {3589},
  year    = {2022},
  doi     = {10.3390/electronics11213589}
}

@article{Das2021_2Dreview,
  author  = {Das, S. and Sebastian, A. and Pop, E. and
             McClellan, C. J. and Franklin, A. D. and
             Grasser, T. and Knobloch, T. and Illarionov, Y. and
             Penumatcha, A. V. and Appenzeller, J. and
             Chen, Z. and Zhu, W. and Asselberghs, I. and
             Li, L.-J. and Avci, U. E. and Bhat, N. and
             Anthopoulos, T. D. and Singh, R.},
  title   = {Transistors Based on Two-Dimensional Materials
             for Future Integrated Circuits},
  journal = {Nat. Electron.},
  volume  = {4},
  pages   = {786--799},
  year    = {2021},
  doi     = {10.1038/s41928-021-00670-1}
}

@article{Fiori2014_2Delectronics,
  author  = {Fiori, G. and Bonaccorso, F. and Iannaccone, G. and
             Palacios, T. and Neumaier, D. and Seabaugh, A. and
             Banerjee, S. K. and Colombo, L.},
  title   = {Electronics Based on Two-Dimensional Materials},
  journal = {Nat. Nanotechnol.},
  volume  = {9},
  pages   = {768--779},
  year    = {2014},
  doi     = {10.1038/nnano.2014.207}
}

@article{Liu2012_ALD,
  author  = {Liu, H. and Xu, K. and Zhang, X. and Ye, P. D.},
  title   = {The Integration of High-k Dielectric on
             Two-Dimensional Crystals by Atomic Layer Deposition},
  journal = {Appl. Phys. Lett.},
  volume  = {100},
  pages   = {152115},
  year    = {2012},
  doi     = {10.1063/1.3703595}
}

@article{Arutchelvan2023_process,
  author  = {O'Brien, K. P. and Naylor, C. H. and Dorow, C. and
             Maxey, K. and Penumatcha, A. V. and Vyatskikh, A. and
             Zhong, T. and Kitamura, A. and Lee, S. and Rogan, C.
             and Mortelmans, W. and Kavrik, M. S. and
             Steinhardt, R. and Buragohain, P. and Dutta, S. and
             Tronic, T. and Clendenning, S. and Fischer, P. and
             Putna, E. S. and Radosavljevic, M. and Metz, M. and
             Avci, U.},
  title   = {Process Integration and Future Outlook of {2D}
             Transistors},
  journal = {Nat. Commun.},
  volume  = {14},
  pages   = {6400},
  year    = {2023},
  doi     = {10.1038/s41467-023-41779-5}
}

@article{Geim2013_vdW,
  author  = {Geim, A. K. and Grigorieva, I. V.},
  title   = {Van der {W}aals Heterostructures},
  journal = {Nature},
  volume  = {499},
  pages   = {419--425},
  year    = {2013},
  doi     = {10.1038/nature12385}
}

@article{Allain2014_WSe2mobility,
  author  = {Allain, A. and Kis, A.},
  title   = {Electron and Hole Mobilities in Single-Layer {WSe$_2$}},
  journal = {ACS Nano},
  volume  = {8},
  pages   = {7180--7185},
  year    = {2014},
  doi     = {10.1021/nn5021538}
}

@article{Movva2015_WSe2dualgate,
  author  = {Movva, H. C. P. and Rai, A. and Kang, S. and
             Kim, K. and Fallahazad, B. and Taniguchi, T. and
             Watanabe, K. and Tutuc, E. and Banerjee, S. K.},
  title   = {High-Mobility Holes in Dual-Gated {WSe$_2$}
             Field-Effect Transistors},
  journal = {ACS Nano},
  volume  = {9},
  pages   = {10402--10410},
  year    = {2015},
  doi     = {10.1021/acsnano.5b04611}
}

@article{Movva2018,
  author  = {Movva, H. C. P. and Lovorn, T. and Fallahazad, B. and
             Larentis, S. and Kim, K. and Taniguchi, T. and
             Watanabe, K. and Banerjee, S. K. and MacDonald, A. H.
             and Tutuc, E.},
  title   = {Tunable $\Gamma$--$K$ Valley Populations in Hole-Doped
             Trilayer {WSe}$_2$},
  journal = {Phys. Rev. Lett.},
  volume  = {120},
  pages   = {107703},
  year    = {2018},
  doi     = {10.1103/PhysRevLett.120.107703}
}

@article{Fallahazad2016,
  author  = {Fallahazad, B. and Movva, H. C. P. and Kim, K. and
             Larentis, S. and Taniguchi, T. and Watanabe, K. and
             Banerjee, S. K. and Tutuc, E.},
  title   = {Shubnikov--de Haas Oscillations of High-Mobility Holes in
             Monolayer and Bilayer {WSe}$_2$: Landau Level Degeneracy,
             Effective Mass, and Negative Compressibility},
  journal = {Phys. Rev. Lett.},
  volume  = {116},
  pages   = {086601},
  year    = {2016},
  doi     = {10.1103/PhysRevLett.116.086601}
}

@article{Xu2017_QHE,
  author  = {Xu, S. and Shen, J. and Long, G. and Wu, Z. and
             Bao, Z. and Liu, C.-C. and Xiao, X. and Han, T. and
             Lin, J. and Wu, Y. and Lu, H. and Hou, J. and An, L. and
             Wang, Y. and Cai, Y. and Ho, K. M. and He, Y. and
             Lortz, R. and Zhang, F. and Wang, N.},
  title   = {Odd-Integer Quantum Hall States and Giant Spin Susceptibility
             in p-Type Few-Layer {WSe}$_2$},
  journal = {Phys. Rev. Lett.},
  volume  = {118},
  pages   = {067702},
  year    = {2017},
  doi     = {10.1103/PhysRevLett.118.067702}
}

@article{Mak2018_valleyreview,
  author  = {Mak, K. F. and Xiao, D. and Shan, J.},
  title   = {Light--Valley Interactions in 2D Semiconductors},
  journal = {Nat. Photon.},
  volume  = {12},
  pages   = {451--460},
  year    = {2018},
  doi     = {10.1038/s41566-018-0204-6}
}

@article{Xiao2007_valleyHall,
  author  = {Xiao, D. and Yao, W. and Niu, Q.},
  title   = {Valley-Contrasting Physics in Graphene: Magnetic Moment
             and Topological Transport},
  journal = {Phys. Rev. Lett.},
  volume  = {99},
  pages   = {236809},
  year    = {2007},
  doi     = {10.1103/PhysRevLett.99.236809}
}

@article{Mak2014_valleyHall,
  author  = {Mak, K. F. and McGill, K. L. and Park, J. and McEuen, P. L.},
  title   = {The Valley {Hall} Effect in {MoS}$_2$ Transistors},
  journal = {Science},
  volume  = {344},
  pages   = {1489--1492},
  year    = {2014},
  doi     = {10.1126/science.1250140}
}

@article{Shen2021_contact,
  author  = {Shen, P.-C. and Su, C. and Lin, Y. and Chou, A.-S. and
             Cheng, C.-C. and Park, J.-H. and Chiu, M.-H. and
             Lu, A.-Y. and Tang, H.-L. and Tavakoli, M. M. and
             Pitner, G. and Ji, X. and Cai, Z. and Mao, N. and
             Wang, J. and Tung, V. and Li, J. and Bokor, J. and
             Zettl, A. and Wu, C.-I. and Palacios, T. and
             Li, L.-J. and Kong, J.},
  title   = {Ultralow Contact Resistance between Semimetal and
             Monolayer Semiconductors},
  journal = {Nature},
  volume  = {593},
  pages   = {211--217},
  year    = {2021},
  doi     = {10.1038/s41586-021-03472-9}
}

@article{Pack2024_WSe2,
  author  = {Pack, J. and Guo, Y. and Liu, Z. and Jessen, B. S. and
             Holtzman, L. and Liu, S. and Cothrine, M. and
             Watanabe, K. and Taniguchi, T. and Mandrus, D. G. and
             Barmak, K. and Hone, J. and Dean, C. R.},
  title   = {Charge-Transfer Contacts for the Measurement of Correlated
             States in High-Mobility {WSe}$_2$},
  journal = {Nat. Nanotechnol.},
  volume  = {19},
  pages   = {948--954},
  year    = {2024},
  doi     = {10.1038/s41565-024-01702-5}
}

@article{Smets2023_WSe2improvements,
  author  = {Patoary, N. H. and Xie, J. and Zhou, G. and
             Al Mamun, F. and Sayyad, M. and Tongay, S. and
             Sanchez Esqueda, I.},
  title   = {Improvements in {2D} p-Type {WSe$_2$} Transistors
             towards Ultimate {CMOS} Scaling},
  journal = {Sci. Rep.},
  volume  = {13},
  pages   = {3304},
  year    = {2023},
  doi     = {10.1038/s41598-023-30317-4}
}

@article{Ali2024_WSe2SS,
  author  = {Ali, F. and Choi, H. and Ali, N. and Hassan, Y. and
             Ngo, T. D. and Ahmed, F. and Park, W. K. and
             Sun, Z. and Yoo, W. J.},
  title   = {Achieving Near-Ideal Subthreshold Swing in p-Type
             {WSe$_2$} Field-Effect Transistors},
  journal = {Adv. Electron. Mater.},
  volume  = {10},
  pages   = {2400071},
  year    = {2024},
  doi     = {10.1002/aelm.202400071}
}

@article{Zhao2013_WSe2layers,
  author  = {Zhao, W. and Ghorannevis, Z. and Chu, L. and
             Toh, M. and Kloc, C. and Tan, P.-H. and Eda, G.},
  title   = {Evolution of Electronic Structure in Atomically
             Thin Sheets of {WS$_2$} and {WSe$_2$}},
  journal = {ACS Nano},
  volume  = {7},
  pages   = {791--797},
  year    = {2013},
  doi     = {10.1021/nn305275h}
}

@article{Desai2014_WSe2strain,
  author  = {Desai, S. B. and Seol, G. and Kang, J. S. and
             Fang, H. and Battaglia, C. and Kapadia, R. and
             Ager, J. W. and Guo, J. and Javey, A.},
  title   = {Strain-Induced Indirect to Direct Bandgap Transition
             in Multilayer {WSe$_2$}},
  journal = {Nano Lett.},
  volume  = {14},
  pages   = {4592--4597},
  year    = {2014},
  doi     = {10.1021/nl501638a}
}

@article{Wickramaratne2014_indirect,
  author  = {Wickramaratne, D. and Zahid, F. and Lake, R. K.},
  title   = {Electronic and Thermoelectric Properties of
             Few-Layer Transition Metal Dichalcogenides},
  journal = {J. Chem. Phys.},
  volume  = {140},
  pages   = {124710},
  year    = {2014},
  doi     = {10.1063/1.4869142}
}

@article{Komsa2012_indirect,
  author  = {Komsa, H.-P. and Krasheninnikov, A. V.},
  title   = {Effects of Confinement and Environment on the
             Electronic Structure and Exciton Binding Energy of
             {MoS$_2$} from First Principles},
  journal = {Phys. Rev. B},
  volume  = {86},
  pages   = {241201},
  year    = {2012},
  doi     = {10.1103/PhysRevB.86.241201}
}

@article{Splendiani2010_MoS2PL,
  author  = {Splendiani, A. and Sun, L. and Zhang, Y. and
             Li, T. and Kim, J. and Chim, C.-Y. and
             Galli, G. and Wang, F.},
  title   = {Emerging Photoluminescence in Monolayer {MoS$_2$}},
  journal = {Nano Lett.},
  volume  = {10},
  pages   = {1271--1275},
  year    = {2010},
  doi     = {10.1021/nl903868w}
}

@article{Mak2010_MoS2direct,
  author  = {Mak, K. F. and Lee, C. and Hone, J. and
             Shan, J. and Heinz, T. F.},
  title   = {Atomically Thin {MoS$_2$}: A New Direct-Gap
             Semiconductor},
  journal = {Phys. Rev. Lett.},
  volume  = {105},
  pages   = {136805},
  year    = {2010},
  doi     = {10.1103/PhysRevLett.105.136805}
}

@article{Yeh2015_WSe2muARPES,
  author  = {Yeh, P.-C. and Jin, W. and Zaki, N. and Zhang, D. and
             Liou, J. T. and Sadowski, J. T. and Al-Mahboob, A. and
             Dadap, J. I. and Herman, I. P. and Sutter, P. and
             Osgood, R. M., Jr.},
  title   = {Layer-Dependent Electronic Structure of an Atomically
             Heavy Two-Dimensional Dichalcogenide},
  journal = {Phys. Rev. B},
  volume  = {91},
  pages   = {041407},
  year    = {2015},
  doi     = {10.1103/PhysRevB.91.041407}
}

@article{Kormanyos2015_kp,
  author  = {Korm{\'a}nyos, A. and Burkard, G. and Gmitra, M. and
             Fabian, J. and Z{\'o}lyomi, V. and Drummond, N. D. and
             Fal'ko, V.},
  title   = {k$\cdot$p Theory for Two-Dimensional Transition Metal
             Dichalcogenide Semiconductors},
  journal = {2D Mater.},
  volume  = {2},
  pages   = {022001},
  year    = {2015},
  doi     = {10.1088/2053-1583/2/2/022001}
}

@article{Gunst2025_WSe2mobility,
  author  = {Ha, V.-A. and Tiwari, S. and Giustino, F.},
  title   = {Ultrahigh Hole Mobility in Monolayer {WSe$_2$}
             Enabled by Spin--Orbit Suppression of Intervalley
             Scattering},
  journal = {Nano Lett.},
  volume  = {25},
  pages   = {14304--14309},
  year    = {2025},
  doi     = {10.1021/acs.nanolett.5c03258}
}

@article{Xiao2012_valley,
  author  = {Xiao, D. and Liu, G.-B. and Feng, W. and
             Xu, X. and Yao, W.},
  title   = {Coupled Spin and Valley Physics in Monolayers of
             {MoS$_2$} and Other Group-{VI} Dichalcogenides},
  journal = {Phys. Rev. Lett.},
  volume  = {108},
  pages   = {196802},
  year    = {2012},
  doi     = {10.1103/PhysRevLett.108.196802}
}

@article{Manzeli2017_TMDreview,
  author  = {Manzeli, S. and Ovchinnikov, D. and Pasquier, D. and
             Yazyev, O. V. and Kis, A.},
  title   = {2{D} Transition Metal Dichalcogenides},
  journal = {Nat. Rev. Mater.},
  volume  = {2},
  pages   = {17033},
  year    = {2017},
  doi     = {10.1038/natrevmats.2017.33}
}

@article{Xu2014_spinpseudo,
  author  = {Xu, X. and Yao, W. and Xiao, D. and Heinz, T. F.},
  title   = {Spin and Pseudospins in Layered Transition Metal
             Dichalcogenides},
  journal = {Nat. Phys.},
  volume  = {10},
  pages   = {343--350},
  year    = {2014},
  doi     = {10.1038/nphys2942}
}

@article{He2024_strainFET,
  author  = {Kumar, A. and Xu, L. and Pal, A. and Agashiwala, K. and
             Parto, K. and Cao, W. and Banerjee, K.},
  title   = {Strain Engineering in {2D} {FETs}: Physics, Status,
             and Prospects},
  journal = {J. Appl. Phys.},
  volume  = {136},
  pages   = {090901},
  year    = {2024},
  doi     = {10.1063/5.0211555}
}

@article{Schmidt2024_straintransfer,
  author  = {Michail, A. and Yang, J. A. and Filintoglou, K. and
             Balakeras, N. and Nattoo, C. A. and Bailey, C. S. and
             Daus, A. and Parthenios, J. and Pop, E. and
             Papagelis, K.},
  title   = {Biaxial Strain Transfer in Monolayer {MoS$_2$}
             and {WSe$_2$} Transistor Structures},
  journal = {ACS Appl. Mater. Interfaces},
  volume  = {16},
  pages   = {49602--49611},
  year    = {2024},
  doi     = {10.1021/acsami.4c07216}
}

@article{Johari2016_WSe2strain,
  author  = {Johari, P. and Shenoy, V. B.},
  title   = {Tuning the Electronic Properties of Semiconducting
             Transition Metal Dichalcogenides by Applying
             Mechanical Strains},
  journal = {ACS Nano},
  volume  = {6},
  pages   = {5449--5456},
  year    = {2012},
  doi     = {10.1021/nn301320r}
}

@article{Lau2024_straintronics,
  author  = {Boland, C. S. and Sun, Y. and Papageorgiou, D. G.},
  title   = {Bandgap Engineering of {2D} Materials toward
             High-Performing Straintronics},
  journal = {Nano Lett.},
  volume  = {24},
  pages   = {12722--12732},
  year    = {2024},
  doi     = {10.1021/acs.nanolett.4c03321}
}

@article{Bennett2023_QC,
  author  = {Bennett, R. K. A. and Pop, E.},
  title   = {How Do Quantum Effects Influence the Capacitance
             and Carrier Density of Monolayer {MoS$_2$}
             Transistors?},
  journal = {Nano Lett.},
  volume  = {23},
  pages   = {1666--1672},
  year    = {2023},
  doi     = {10.1021/acs.nanolett.2c03913}
}

@article{Luryi1988_QC,
  author  = {Luryi, S.},
  title   = {Quantum Capacitance Devices},
  journal = {Appl. Phys. Lett.},
  volume  = {52},
  pages   = {501--503},
  year    = {1988},
  doi     = {10.1063/1.99649}
}

@article{Radisavljevic2011_MoS2,
  author  = {Radisavljevic, B. and Radenovic, A. and Brivio, J.
             and Giacometti, V. and Kis, A.},
  title   = {Single-Layer {MoS$_2$} Transistors},
  journal = {Nat. Nanotechnol.},
  volume  = {6},
  pages   = {147--150},
  year    = {2011},
  doi     = {10.1038/nnano.2010.279}
}

@article{Guo2015_trapping,
  author  = {Guo, Y. and Wei, X. and Shu, J. and Liu, B. and
             Yin, J. and Guan, C. and Han, Y. and Gao, S. and
             Chen, Q.},
  title   = {Charge Trapping at the {MoS$_2$--SiO$_2$} Interface
             and Its Effects on the Characteristics of {MoS$_2$}
             Metal-Oxide-Semiconductor Field Effect Transistors},
  journal = {Appl. Phys. Lett.},
  volume  = {106},
  pages   = {103109},
  year    = {2015},
  doi     = {10.1063/1.4914968}
}

@article{Illarionov2020_Ditreview,
  author  = {Illarionov, Y. Y. and Knobloch, T. and Jech, M. and
             Lanza, M. and Akinwande, D. and Vexler, M. I. and
             Mueller, T. and Lemme, M. C. and Fiori, G. and
             Schwierz, F. and Grasser, T.},
  title   = {Insulators for 2{D} Nanoelectronics: The Gap to
             Bridge},
  journal = {Nat. Commun.},
  volume  = {11},
  pages   = {3385},
  year    = {2020},
  doi     = {10.1038/s41467-020-16640-8}
}

@article{Ghatak2015_trapping,
  author  = {Ghatak, S. and Mukherjee, S. and Jain, M. and
             Sarma, D. D. and Ghosh, A.},
  title   = {Microscopic Origin of Low Frequency Noise in
             {MoS$_2$} Field-Effect Transistors},
  journal = {APL Mater.},
  volume  = {2},
  pages   = {092515},
  year    = {2014},
  doi     = {10.1063/1.4895955}
}

@article{Leonhardt2017_Dit,
  author  = {Gaur, A. and Balaji, Y. and Lin, D. and Adelmann, C. and
             Van Houdt, J. and Heyns, M. and Mocuta, D. and Radu, I.},
  title   = {Demonstration of $2\times10^{12}$~cm$^{-2}$ eV$^{-1}$
             2{D}-Oxide Interface Trap Density on Back-Gated
             {MoS$_2$} Flake Devices with 2.5~nm {EOT}},
  journal = {Microelectron. Eng.},
  volume  = {178},
  pages   = {145--149},
  year    = {2017},
  doi     = {10.1016/j.mee.2017.05.006}
}

@inproceedings{Lee2022IEDM_MoS2SS,
  author    = {Lee, T.-E. and Su, Y.-C. and Lin, B.-J. and
               Chen, Y.-X. and Yun, W.-S. and Ho, P.-H. and
               Wang, J.-F. and Su, S.-K. and Hsu, C.-F. and
               Mao, P.-S. and Chang, Y.-C. and Chien, C.-H. and
               Liu, B.-H. and Su, C.-Y. and Kei, C.-C. and
               Wang, H. and Wong, H.-S. P. and Lee, T. Y. and
               Chang, W.-H. and Cheng, C.-C. and Radu, I. P.},
  title     = {Nearly Ideal Subthreshold Swing in Monolayer
               {MoS$_2$} Top-Gate n{FET}s with Scaled {EOT} of 1\,nm},
  booktitle = {2022 IEEE International Electron Devices Meeting (IEDM)},
  pages     = {7.4.1--7.4.4},
  year      = {2022},
  doi       = {10.1109/IEDM45625.2022.10019552}
}

@article{Cao2014_compactmodel,
  author  = {Cao, W. and Kang, J. and Liu, W. and Banerjee, K.},
  title   = {A Compact Current--Voltage Model for {2D}
             Semiconductor Based Field-Effect Transistors
             Considering Interface Traps, Mobility Degradation,
             and Inefficient Doping Effect},
  journal = {IEEE Trans. Electron Devices},
  volume  = {61},
  pages   = {4282--4290},
  year    = {2014},
  doi     = {10.1109/TED.2014.2365028}
}

@article{Jimenez2012_driftdiff,
  author  = {Jim{\'e}nez, D.},
  title   = {Drift-Diffusion Model for Single Layer Transition
             Metal Dichalcogenide Field-Effect Transistors},
  journal = {Appl. Phys. Lett.},
  volume  = {101},
  pages   = {243501},
  year    = {2012},
  doi     = {10.1063/1.4770313}
}

@article{Zhu2011_SOC,
  author  = {Zhu, Z. Y. and Cheng, Y. C. and
             Schwingenschl{\"o}gl, U.},
  title   = {Giant Spin-Orbit-Induced Spin Splitting in
             Two-Dimensional Transition-Metal Dichalcogenide
             Semiconductors},
  journal = {Phys. Rev. B},
  volume  = {84},
  pages   = {153402},
  year    = {2011},
  doi     = {10.1103/PhysRevB.84.153402}
}

@article{Riley2014_WSe2spinvalley,
  author  = {Riley, J. M. and Mazzola, F. and Dendzik, M. and
             Michiardi, M. and Takayama, T. and Bawden, L. and
             Graner{\o}d, C. and Leandersson, M. and
             Balasubramanian, T. and Hoesch, M. and Kim, T. K. and
             Takagi, H. and Meevasana, W. and Hofmann, Ph. and
             Bahramy, M. S. and Wells, J. W. and King, P. D. C.},
  title   = {Direct Observation of Spin-Polarized Bulk Bands in
             an Inversion-Symmetric Semiconductor},
  journal = {Nat. Phys.},
  volume  = {10},
  pages   = {835--839},
  year    = {2014},
  doi     = {10.1038/nphys3105}
}

@article{Tang2025_highk,
  author  = {Kang, T. and Park, J. and Lee, S. Y. and Jung, H.
             and Jeon, J. and Kim, Y.-H. and Lee, S.},
  title   = {High-$\kappa$ Dielectric van der {W}aals Integration
             on {2D} Semiconductors for Three-Dimensional
             Complementary Logic Systems},
  journal = {Nat. Commun.},
  volume  = {16},
  pages   = {11648},
  year    = {2025},
  doi     = {10.1038/s41467-025-66770-0}
}

@article{Radisavljevic2012_dualgate,
  author  = {Radisavljevic, B. and Whitwick, M. B. and Kis, A.},
  title   = {Integrated Circuits and Logic Operations Based on
             Single-Layer {MoS$_2$}},
  journal = {ACS Nano},
  volume  = {5},
  pages   = {9934--9938},
  year    = {2011},
  doi     = {10.1021/nn203715c}
}

@article{Novoselov2016_2Dscience,
  author  = {Novoselov, K. S. and Mishchenko, A. and Carvalho, A.
             and Castro Neto, A. H.},
  title   = {{2D} Materials and Van der {W}aals Heterostructures},
  journal = {Science},
  volume  = {353},
  pages   = {aac9439},
  year    = {2016},
  doi     = {10.1126/science.aac9439}
}

@article{Schaibley2016_valleytronics,
  author  = {Schaibley, J. R. and Yu, H. and Clark, G. and
             Rivera, P. and Ross, J. S. and Seyler, K. L. and
             Yao, W. and Xu, X.},
  title   = {Valleytronics in {2D} Materials},
  journal = {Nat. Rev. Mater.},
  volume  = {1},
  pages   = {16055},
  year    = {2016},
  doi     = {10.1038/natrevmats.2016.55}
}

@article{Sohier2018_mobility,
  author  = {Sohier, T. and Campi, D. and Marzari, N. and Gibertini, M.},
  title   = {Mobility of Two-Dimensional Materials from First Principles
             in an Accurate and Automated Framework},
  journal = {Phys. Rev. Mater.},
  volume  = {2},
  pages   = {114010},
  year    = {2018},
  doi     = {10.1103/PhysRevMaterials.2.114010}
}

@article{Sohier2019_valleyEng,
  author  = {Sohier, T. and Gibertini, M. and Campi, D. and
             Pizzi, G. and Marzari, N.},
  title   = {Valley-Engineering Mobilities in Two-Dimensional Materials},
  journal = {Nano Lett.},
  volume  = {19},
  pages   = {3723--3729},
  year    = {2019},
  doi     = {10.1021/acs.nanolett.9b00865}
}

@article{Afrid2026_intervalley,
  author  = {Ta-Seen Afrid, Sheikh Mohd and Zhao, He Lin and
             van der Zande, Arend M. and Rakheja, Shaloo},
  title   = {Strain-Tunable Inter-Valley Scattering Defines
             Universal Mobility Enhancement in n- and p-Type
             {2D} {TMDs}},
  journal = {npj 2D Mater. Appl.},
  volume  = {10},
  pages   = {57},
  year    = {2026},
  doi     = {10.1038/s41699-026-00689-y}
}

@article{Wang2012,
  author  = {Wang, Q. H. and Kalantar-Zadeh, K. and Kis, A. and Coleman, J. N. and Strano, M. S.},
  title   = {Electronics and Optoelectronics of Two-Dimensional Transition Metal Dichalcogenides},
  journal = {Nat. Nanotechnol.},
  volume  = {7},
  pages   = {699--712},
  year    = {2012},
  doi     = {10.1038/nnano.2012.193}
}

@article{Desai2016,
  author  = {Desai, S. B. and Madhvapathy, S. R. and Sachid, A. B. and
             Llinas, J. P. and Wang, Q. and Ahn, G. H. and Pitner, G. and
             Kim, M. J. and Bokor, J. and Hu, C. and Wong, H.-S. P. and Javey, A.},
  title   = {{MoS$_2$} Transistors with 1-Nanometer Gate Lengths},
  journal = {Science},
  volume  = {354},
  pages   = {99--102},
  year    = {2016},
  doi     = {10.1126/science.aah4698}
}

@article{Liu2021nature,
  author  = {Liu, Y. and Duan, X. and Shin, H.-J. and Park, S. and Huang, Y. and Duan, X.},
  title   = {Promises and prospects of two-dimensional transistors},
  journal = {Nature},
  volume  = {591},
  pages   = {43--53},
  year    = {2021},
  doi     = {10.1038/s41586-021-03339-z}
}

@article{Fivaz1967,
  author  = {Fivaz, R. and Mooser, E.},
  title   = {Mobility of Charge Carriers in Semiconducting Layer Structures},
  journal = {Phys. Rev.},
  volume  = {163},
  pages   = {743--755},
  year    = {1967},
  doi     = {10.1103/PhysRev.163.743}
}

@article{Allain2015,
  author  = {Allain, A. and Kang, J. and Banerjee, K. and Kis, A.},
  title   = {Electrical Contacts to Two-Dimensional Semiconductors},
  journal = {Nat. Mater.},
  volume  = {14},
  pages   = {1195--1205},
  year    = {2015},
  doi     = {10.1038/nmat4452}
}

@article{Ramasubramaniam2011,
  author  = {Ramasubramaniam, A. and Naveh, D. and Towe, E.},
  title   = {Tunable band gaps in bilayer transition-metal dichalcogenides},
  journal = {Phys. Rev. B},
  volume  = {84},
  pages   = {205325},
  year    = {2011},
  doi     = {10.1103/PhysRevB.84.205325}
}

@article{Zhang2014_MoSe2,
  author  = {Zhang, Y. and Chang, T.-R. and Zhou, B. and Cui, Y.-T. and
             Yan, H. and Liu, Z. and Schmitt, F. and Lee, J. and Moore, R. and
             Chen, Y. and Lin, H. and Jeng, H.-T. and Mo, S.-K. and
             Hussain, Z. and Bansil, A. and Shen, Z.-X.},
  title   = {Direct Observation of the Transition from Indirect to Direct
             Bandgap in Atomically Thin Epitaxial {MoSe$_2$}},
  journal = {Nat. Nanotechnol.},
  volume  = {9},
  pages   = {111--115},
  year    = {2014},
  doi     = {10.1038/nnano.2013.277}
}

@article{Liang2023CPL,
  author  = {Liang, B. and Liu, L. and Tang, J. and Chen, J. and Shi, Y. and Li, S.},
  title   = {Enhancement of Carrier Mobility in Semiconductor Nanostructures
             by Carrier Distribution Engineering},
  journal = {Chin. Phys. Lett.},
  volume  = {40},
  pages   = {058503},
  year    = {2023},
  doi     = {10.1088/0256-307X/40/5/058503}
}

@article{Ahn2017NatComm,
  author  = {Ahn, G. H. and Amani, M. and Rasool, H. and Lien, D.-H. and
             Mastandrea, J. P. and Ager, J. W. and Dubey, M. and
             Chrzan, D. C. and Minor, A. M. and Javey, A.},
  title   = {Strain-Engineered Growth of Two-Dimensional Materials},
  journal = {Nat. Commun.},
  volume  = {8},
  pages   = {608},
  year    = {2017},
  doi     = {10.1038/s41467-017-00516-5}
}

@article{Lin2025NatElectron,
  author  = {Lin, C.-Y. and Chen, B.-C. and Liu, Y.-C. and Kuo, S.-F. and
             Tsai, H.-C. and Chang, Y.-M. and Kuo, C.-Y. and Chang, C.-F. and
             Chen, J.-H. and Chu, Y.-H. and Yamamoto, M. and Shen, C.-H. and
             Chueh, Y.-L. and Chiu, P.-W. and Chen, Y.-C. and Yang, J.-C. and
             Lin, Y.-F.},
  title   = {Integration of Freestanding Hafnium Zirconium Oxide Membranes
             into Two-Dimensional Transistors as a High-$\kappa$ Ferroelectric
             Dielectric},
  journal = {Nat. Electron.},
  volume  = {8},
  pages   = {560--570},
  year    = {2025},
  doi     = {10.1038/s41928-025-01398-y}
}

@article{Ghosh2025_bilayerWSe2FET,
  author  = {Ghosh, S. and Sadaf, M. U. K. and Graves, A. R. and
             Zheng, Y. and Pannone, A. and Ray, S. and Cheng, C.-Y. and
             Guevara, J. and Redwing, J. M. and Das, S.},
  title   = {High-Performance p-Type Bilayer {WSe$_2$} Field Effect
             Transistors by Nitric Oxide Doping},
  journal = {Nat. Commun.},
  volume  = {16},
  pages   = {5649},
  year    = {2025},
  doi     = {10.1038/s41467-025-59684-4}
}

@article{Tang2025_GAA2D,
  author  = {Tang, J. and Jiang, J. and Gao, X. and Gao, X. and
             Zhang, C. and Wang, M. and Xue, C. and Li, Z. and
             Yin, Y. and Tan, C. and Ding, F. and Qiu, C. and
             Peng, L.-M. and Peng, H.},
  title   = {Low-Power {2D} Gate-All-Around Logics via Epitaxial
             Monolithic {3D} Integration},
  journal = {Nat. Mater.},
  volume  = {24},
  pages   = {519--526},
  year    = {2025},
  doi     = {10.1038/s41563-025-02117-w}
}

@article{Godiksen2022_valley,
  author  = {Godiksen, R. H. and Wang, S. and Raziman, T. V. and
             Rivas, J. G. and Curto, A. G.},
  title   = {Impact of Indirect Transitions on Valley Polarization
             in {WS$_2$} and {WSe$_2$}},
  journal = {Nanoscale},
  volume  = {14},
  pages   = {17761--17769},
  year    = {2022},
  doi     = {10.1039/D2NR04800K}
}

@article{Lai2023_valleytronic,
  author  = {Lai, S. and Zhang, Z. and Wang, N. and Rasmita, A. and
             Deng, Y. and Liu, Z. and Gao, W.-B.},
  title   = {Dual-Gate All-Electrical Valleytronic Transistors},
  journal = {Nano Lett.},
  volume  = {23},
  pages   = {192--197},
  year    = {2023},
  doi     = {10.1021/acs.nanolett.2c03947}
}

@article{Li2022_ferroelectric_valley,
  author  = {Li, X. and Yang, C. and Xia, Y. and Zeng, X. and
             Shen, P. and Li, L. and Xu, F. and Cai, D. and
             Wu, Y. and Wu, Z. and Li, S. and Kang, J.},
  title   = {Nonvolatile Electrical Valley Manipulation in {WS$_2$}
             by Ferroelectric Gating},
  journal = {ACS Nano},
  volume  = {16},
  pages   = {20598--20606},
  year    = {2022},
  doi     = {10.1021/acsnano.2c07469}
}

@article{Bae2022_intervalley,
  author  = {Bae, S. and Matsumoto, K. and Raebiger, H. and
             Shudo, K. and Kim, Y.-H. and
             Handeg{\aa}rd, {\O}. S. and
             Nagao, T. and Kitajima, M. and Sakai, Y. and
             Zhang, X. and Vajtai, R. and Ajayan, P. and
             Kono, J. and Takeda, J. and Katayama, I.},
  title   = {K-Point Longitudinal Acoustic Phonons Are Responsible
             for Ultrafast Intervalley Scattering in Monolayer {MoSe$_2$}},
  journal = {Nat. Commun.},
  volume  = {13},
  pages   = {4279},
  year    = {2022},
  doi     = {10.1038/s41467-022-32008-6}
}

\end{document}